\documentclass{article}
\usepackage{amsfonts}
\usepackage{graphicx}
\usepackage{float}
\usepackage{color}
\usepackage{amsmath,amssymb}
\usepackage{verbatim}

\begin{document}

\title{Perturbed Kepler problem in general relativity with quaternions}
\author{ 
F. Nemes \\
ELTE Department of Atomic Physics \\
1117 Budapest P\'azm\'any P\'eter street 1/A \\
and \\
B. Mik\'oczi \\
MTA Wigner FK, Research Institute for Particle and Nuclear Physics \\
Budapest 114, P.O.~Box 49, H-1525, Hungary }
\maketitle

\begin{abstract}

The motion of binary star systems is re-examined in the presence of perturbations from the 
theory of general relativity. To handle the singularity of the Kepler problem, the equation
of motion is regularized and linearized with quaternions. In this way first order perturbation 
results are derived using the quaternion based approach. 

\end{abstract}

\pagestyle{plain}
\section{Introduction}
In this paper gravitational effects as perturbations of the
Kepler problem are examined with post-Newtonian approximation. Gravitational effects become strong 
when the components of the binary are close to each other, and the orbital separation is small.\footnote{The post-Newtonian 
approximation is not valid when the orbital separation is smaller than the innermost stable circular orbit or when 
the bodies start to merge. Therefore in the paper it is supposed that these limits are not reached.} The Kepler problem 
is singular when the separation is zero, therefore to study gravitational effects the desingularization -- or regularization --
of the equation of motion would be a substantial step.

It is well known that Kustaanheimo (1964) solved the regularization of the three-dimensional Kepler problem with spinors\cite{Kustaanheimo}, which 
was reformulated by Stiefel\cite{Stiefel}. In their method -- the KS method for short -- the regularization 
was carried out in four dimensions, and it was proved that the three-dimensional Kepler problem can only be 
regularized using four-dimensional linear spaces.

We follow another approach developed by J. Vrbik. In his work the mentioned four-dimensional 
space is the linear space of quaternions and the regularization is calculated with quaternion algebra. He applied 
his method with success to the Lunar problem\cite{VrbikLunar}, and several perturbative forces were 
studied in details\cite{Vrbikbook,Vrbik1,VrbikJMP}.

In the present work we use his method to examine gravitational effects analytically with quaternions. The 
leading order correction of general relativity to classical mechanics is calculated first. 
The formula for the precession of the pericentre is derived based on the Vrbik's quaternion formulae. Then 
the gravitational radiation reaction is analyzed, where the famous Peters-Mathews formula is 
proved\cite{Peters:1963ux}. In this calculation we manage to remove the residual coordinate gauge freedom of 
the gravitational reaction from the quaternionic equation of motion. In addition using a one-dimensional
model we demonstrate that the regularization can lead to different results depending on that the Sundman 
transformation is employed with the perturbed or unperturbed orbital separation. 

The regularization is defined with four-dimensional spaces, thus an additional geometrical constraint 
-- a gauge -- have to be applied to describe the three-dimensional spatial Kepler problem. In the KS method 
the so-called \textit{bilinear relation} is defined, which is an excellent gauge for \textit{numerical} calculations. 
Vrbik proposes another constraint to provide an \textit{analytic} perturbative method, since -- according to Vrbik -- the 
bilinear relation is too restrictive to build an analytic perturbative method. This constraint is the major difference between the KS 
method and Vrbik's work.  

The Laplace vector is a constant of motion of the Kepler problem, which is a consequence of the hidden symmetry
of the problem\cite{Landau I}. This symmetry becomes manifest in four dimensions, which shows that the Kepler
problem has another interesting connection with the four-dimensional space.  This connection has far reaching
consequences\cite{Gyoergyi:1969sh,Cordani}.

Quaternions were first applied to regularize the Kepler problem by Chelnokov who successfully regularized
the Kepler problem to describe rotating coordinate systems\cite{Chelnokov1981}. Moreover he was able 
to apply his results to describe the optimal control problem of a spacecraft\cite{Chelnokov2001}.

Later it was shown by Vivarelli (1983) in a general mathematical sense that the KS method can be transformed
identically into quaternion algebra\cite{Vivarelli}. Quaternion algebra proved to be very useful to derive 
the central ideas of the KS method. Remarkably the bilinear relation is described as a fibration of 
the quaternion space.

More recently Waldvogel showed that the spatial Kepler motion can be elegantly formulated with quaternions 
using a novel star conjugation operator\cite{Waldvogel}. The star conjugation is especially useful to handle 
the bilinear relation. The interesting connection with the Birkhoff transformation is also shown \cite{Waldvogel2008}.
Quaternions turned to be useful in case of three and N-body applications\cite{Saha}. 

It has to be emphasized that the mentioned quaternion approaches \textit{exclusively} apply the ``bilinear relation" as 
a gauge to reduce the dimensions from four to three, while Vrbik apply his special gauge.   

This paper is organized as follows: a short outline of Vrbik's approach is provided in 
Section \ref{regularization with quaternion} and \ref{Orbitalelementswithquaternions}, where
we describe the transformation of the Kepler problem into quaternion differential equation. Then 
the solution is given in terms of ordinary differential equations of orbital elements. The advantages of Vrbik's 
calculus compared with the KS method are highlighted.   

In Section~\ref{examplefortheregularization} a one-dimensional example is given where we demonstrate
that the result of the regularization depends on whether the Sundman transformation is applied with
perturbed or unperturbed orbital separation.

In Section~\ref{GRperturbations} Vrbik's method is applied to two perturbations.  First of all, the
leading order correction of general relativity to classical mechanics is examined. The formula for 
the precession of the pericentre is derived. Then the gravitational radiation reaction is analyzed, 
where the famous Peters-Mathews formula is proved using the quaternion approach \cite{Peters:1963ux}. 
In this calculation we solved to cancel the residual coordinate gauge freedom of the gravitational radiation reaction 
in the quaternionic equation of motion.  

The conclusion and the outlook is given in Section~\ref{Conclusion} followed by the Appendix.

\section{Linearization and regularization with quaternions}
\label{regularization with quaternion}
\subsection{Quaternion algebra basics}
The quaternion algebra has three imaginary units, generally called $\mathfrak{i}$, $\mathfrak{j}$ and 
$\mathfrak{k}$, where $\mathfrak{i}^2=\mathfrak{j}^2=\mathfrak{k}^2=-1$. Any of them anticommute 
\begin{equation}
	\mathfrak{i}\mathfrak{j}=-\mathfrak{j}\mathfrak{i},\,
	\mathfrak{j}\mathfrak{k}=-\mathfrak{k}\mathfrak{j},\,
	\mathfrak{k}\mathfrak{i}=-\mathfrak{i}\mathfrak{k}\,,
\end{equation}   
while the real unit 1 commutes with each of them. The four units together form the algebra's 
generators. Thus any element of the algebra can be written as\footnote{%
Note the unusual reversed ordering of the $U_i$ components.}
\begin{equation}
	\mathbb{A}=A+A_{3}\mathfrak{i}+A_{2}\mathfrak{j}+A_{1}\mathfrak{k}=A+\mathbf{a}\,. 
\end{equation} 
Quaternion \textit{conjugation} reverses the sign of the imaginary units  
\begin{equation}
	\mathbb{\bar{A}}=A-\mathbf{a}\,. 
\end{equation}
The \textit{magnitude} of a quaternion is defined as
\begin{equation}
	\left|\mathbb{A}\right|=\sqrt{\bar{\mathbb{A}}\mathbb{A}}\,.
\end{equation} 

Any quaternion can be written in the form $A+a \hat{\mathbf{a}}$, where $a$ is the magnitude and $\hat{\mathbf{a}}$
is the unit direction of $\mathbf{a}$. Since $\hat{\mathbf{a}}^2=-1$ the exponential on any quaternion 
can be expressed with Euler's formula  
\begin{equation}
	e^{A+a\hat{\mathbf{a}}}=e^{A}\left( \cos a + \hat{\mathbf{a}}\sin a \right)\,.
\end{equation} 
\subsubsection{Representation of spatial vectors and rotations}
Spatial vectors are represented with \textit{pure} quaternions, which has no real part
\begin{equation}
	\mathbf{x}=z\mathfrak{i}+y\mathfrak{j}+x\mathfrak{k}\,, 
	\label{unitsaxis}
\end{equation}
where the $z$-axis is associated with the $\mathfrak{i}$ unit. With this interpretation 
quaternion multiplication can be expressed as
\begin{equation}
        \mathbb{AB}=(A+\mathbf{a})(B+\mathbf{b})=AB-\mathbf{a}\cdot\mathbf{b}+A\mathbf{b}+B\mathbf{a}-
        \mathbf{a}\times\mathbf{b}. 
	\label{quaternionmul}
\end{equation}
By substituting $A=B=0$ into this expression~(\ref{quaternionmul}), the anticommutative cross product can
be expressed as 
\begin{equation}
        \mathbf{a}\times\mathbf{b}=-\frac{\mathbf{a}\mathbf{b}-\mathbf{b}\mathbf{a}}{2}. 
	\label{crossproduct}
\end{equation}
Let us introduce a vector $\mathbf{w}$. It is straightforward to show that a rotation 
around the vector $\mathbf{w}$ with magnitude $\left|\mathbf{w}\right|$ can be written 
as\cite{Vrbikbook}
\begin{equation}
	\tilde{\mathbf{x}}=\mathbb{\bar{R}}\mathbf{x}\mathbb{R},
	\label{rotationsingeneral}
\end{equation}
where $\mathbb{R}=e^{\frac{\mathbf{w}}{2}}$. Note that $\mathbb{R\bar{R}}=1$, hence
$\mathbf{x}=\mathbb{R}\tilde{\mathbf{x}}\bar{\mathbb{R}}$ is the inverse rotation. Let us suppose that 
the rotation is parameter dependent $\mathbb{R}(s)$, and differentiate it with respect to $s$
\begin{equation}
        \tilde{\mathbf{x}}'=\mathbb{\bar{R}}\mathbf{x}\mathbb{R}'+\mathbb{\bar{R}}'\mathbf{x}\mathbb{R}=
	\tilde{\mathbf{x}}\bar{\mathbb{R}}\mathbb{R}'+\mathbb{\bar{R}}'\mathbb{R}\tilde{\mathbf{x}}\,,
	\label{rotatitonderivative}
\end{equation}
where definition (\ref{rotationsingeneral}) of $\tilde{\mathbf{x}}$ was applied. With the help of the 
identity $(\mathbb{\bar{R}R})'=\mathbb{\bar{R}'R}+\mathbb{\bar{R}R'}=0$ and the 
cross product (\ref{crossproduct}) this can be further written  
\begin{equation} 
        \tilde{\mathbf{x}}'=
        \tilde{\mathbf{x}}\bar{\mathbb{R}}\mathbb{R}'-\mathbb{\bar{R}}\mathbb{R}'\tilde{\mathbf{x}}=
	\mathbb{Z}\times\tilde{\mathbf{x}},
\end{equation}
where $\mathbb{Z}=2\bar{\mathbb{R}}\mathbb{R}'$. Consequently $\mathbb{Z}$ is the instantaneous angular
velocity of $\tilde{\mathbf{x}}$ with respect to $s$. With an inverse rotation the 
angular velocity $\mathbb{Z}$ can be expressed in a special coordinate system 
-- in the \textit{Kepler frame} -- where $\tilde{\mathbf{x}}$ is instantaneously at rest
\begin{equation}
	\mathbb{Z}_K=\mathbb{R}\mathbb{Z}\bar{\mathbb{R}}=2\mathbb{R}'\bar{\mathbb{R}}\,.
	\label{angularvelocityKepler}
\end{equation}
where the subscript indicates the Kepler frame. 

\subsection{The Kepler problem with quaternions}
Equipped with the quaternion formulae we turn to regularize the Kepler problem with quaternions.
The perturbed Kepler problem in the $G=c=1$ system is given by the equation
\begin{equation}
	{\ddot{\mathbf{r}}}+\frac{m }{r^{3}}\mathbf{r}=\varepsilon \mathbf{f},
	\label{equation of perturbed motion}
\end{equation}%
where $\mathbf{r}$ is the orbital separation vector of the orbiting bodies
\begin{equation}
	\mathbf{r}=\mathbf{r}_{1}-\mathbf{r}_{2},\,r=\left\vert \mathbf{r}%
	\right\vert ,  \label{separation}
\end{equation}%
$\varepsilon $ is the small parameter of the perturbation, and $m =m_{1}+m_{2}$
is the sum of the two masses. 

To regularize the Kepler problem the separation is defined by the
following \textit{quaternionic} equation
\begin{equation}
	\mathbf{r}=\mathbb{\bar{U}}\mathfrak{k}\mathbb{U},
	\label{transformation of the location}
\end{equation}%
where $\mathbb{U}=U+U_{3}\mathfrak{i}+U_{2}\mathfrak{j}+U_{1}\mathfrak{k}$
is a general quaternion. The conjugate of $\mathbf{r}$ is 
\begin{equation}
\bar{\mathbf{r}}=\overline{\mathbb{\bar{U}}\mathfrak{k}\mathbb{U}}=\mathbb{\bar{U}}\mathfrak{\bar{k}}\mathbb{\bar{\bar{U}}}=-\mathbb{\bar{U}}\mathfrak{k}\mathbb{U}=-\mathbf{r},
\label{conjugateofr}
\end{equation}
where the $\overline{\mathbb{A}\mathbb{B}}=\mathbb{\bar{B}}\mathbb{\bar{A}}$, $\bar{\bar{\mathbb{A}}}=\mathbb{A}$  and $\mathfrak{\bar{k}}=-\mathfrak{k}$ properties were used. Therefore
$\mathbf{r}$ has no real part and can be written as $\mathbf{r}=z\mathfrak{i}+y\mathfrak{j}+x\mathfrak{k}$. A direct 
 calculation from (\ref{transformation of the location}) tells us that
\begin{align}
x &= U^{2} + U_{1}^{2}-U_{2}^{2}-U_{3}^{2}\,, \notag \\ 
y &= 2\left( U_{1}U_{2}+U_{3}U\right)\,, \label{KStransformation} \notag \\
z &= 2\left( U_{1}U_{3}-U_{2}U\right) \,, 
\end{align}
and the real part $UU_{1}-U_{3}U_{2}+U_{2}U_{3}-U_{1}U=0$ indeed vanishes. The obtained transformation (\ref{KStransformation}) is just the KS transformation with the $U \rightarrow -U$ convention. \cite{Stiefel}

Transformation (\ref{transformation of the location}) maps the four-dimensional quaternion space into the three-dimensional 
space of spatial vectors. Therefore the solution to a given three-dimensional $\mathbf{r}$
in terms of four-dimensional $\mathbb{U}$ is not unique. From (\ref%
{transformation of the location}) it is clear that the transformation
\begin{equation}
\mathbb{U}\rightarrow e^{\mathfrak{k}\alpha }\mathbb{U},
\label{gauge transformation}
\end{equation}%
where $\alpha$ is an arbitrary real number, is a continuous symmetry of (\ref%
{transformation of the location}). It follows that we have a one-dimensional
compact manifold -- a fibre -- of $\mathbb{U}$s for a given $\mathbf{r}$,
and (\ref{gauge transformation}) defines a fibration of the space of $%
\mathbb{U}$s. The geometrical background of this transformation is elegantly described
in Waldvogel (2005) \cite{Waldvogel}. This additional degree of freedom will be constrained in a
careful manner.

To complete the regularization the time has to be also transformed. The
\textit{Sundman transformation} is given by

\begin{equation}
\frac{dt}{ds}=2r\sqrt{\frac{a}{m}},  \label{transformation of the time}
\end{equation}%
where $s$ is the modified time and $a$ is a real and at this
point \textit{arbitrary} function  of $s$ (it will be chosen such
that it simplifies the solution). From now the operator $^{\prime }$ indicates 
differentiation with respect to the modified time $s$.

Inserting the definition of the orbital separation (\ref{transformation of the
location}) into the equation of motion (\ref{equation of perturbed motion})
while transforming the original time into the modified one using (\ref%
{transformation of the time}) lead us to the following quaternionic
differential equation \cite{Vrbikbook}

\begin{equation}
2\mathbb{U}^{\prime \prime }-\left( 2\mathbb{U}^{\prime }\overline{\mathbb{U}%
^{\prime }}-4a\right) \frac{\mathbb{U}}{r}+2\mathfrak{k}\mathbb{U}^{\prime }%
\frac{\Gamma }{r}+\mathfrak{k}\mathbb{U}\left( \frac{\Gamma }{r}\right)
^{\prime }-\left( \mathbb{U}^{\prime }+\mathfrak{k}\mathbb{U}\frac{\Gamma }{%
2r}\right) \frac{a^{\prime }}{a}+4\frac{a}{m }\varepsilon \mathbb{U}%
\mathbf{rf}=0\,,  \label{regularized equation of motion}
\end{equation}%
where
\begin{equation}
\Gamma =\overline{\mathbb{U}}\mathfrak{k}\mathbb{U}^{\prime }\mathbb{-}%
\overline{\mathbb{U}}^{\prime }\mathfrak{k}\mathbb{U}=2\left( U_{1}U^{\prime
}-UU_{1}^{\prime }+U_{2}U_{3}^{\prime }-U_{3}U_{2}^{\prime }\right) ,
\label{gamma}
\end{equation}%
which is a scalar in the sense that it is invariant under conjugation. This
quantity is the four-dimensional scalar product of the
tangent vector of the $\mathbb{U}\left( s\right) $ curve and the tangent of
the fibre at that point multiplied by two\footnote{
This statement can be checked by differentiating the transformed $\mathbb{U}$ in ($%
\ref{gauge transformation}$) with respect to $\alpha $ which produces the
tangent of the fiber and then taking the four-dimensional scalar product
with $\mathbb{U}^{\prime }$.}. Let us define a condition
\begin{equation}
\Gamma=0\,,  \label{KScondition}
\end{equation}%
which means that the trajectory $\mathbb{U}\left( s\right)$ intersects the fibres 
under right angles. It is indeed a condition as transforming $\mathbb{U}$ with symmetry
transformation (\ref{gauge transformation}) $\Gamma$ transforms as
\begin{equation}
\Gamma \rightarrow \Gamma -2\alpha ^{\prime }r,  \label{gamma trafo}
\end{equation}
hence condition (\ref{KScondition}) can be satisfied by solving a first
order differential equation for $\alpha(s)$.

The geometrical constraint~(\ref{KScondition}) is equivalent with the so-called ``\textit{bilinear relation}",
 which plays an essential role in the KS method\cite{Stiefel}. It can be 
proved that~(\ref{gauge transformation}) is a
dynamical symmetry, since the transformed $\mathbb{U%
}$ solves the equation of motion. Furthermore if condition (\ref{KScondition}) is
satisfied then $\Gamma ^{\prime }$ also vanishes, which means that one can
maintain this condition by finding the proper initial conditions $\mathbb{U}(0)$ and
$\mathbb{U}'(0)$ which satisfy the ``bilinear relation"~(\ref{KScondition})\cite{Vrbikbook}. 

\subsection{Solving the unperturbed case: Kepler orbits}
The unperturbed situation with condition (\ref%
{KScondition}) reduces the perturbed equation of motion (\ref{regularized equation of motion}%
) to the following equation
\begin{equation}
\mathbb{U}^{\prime \prime }-\left(\mathbb{U}^{\prime }\mathbb{\bar{U}}
^{\prime }-2a\right) \frac{\mathbb{U}}{r}=0\,.  
\end{equation}
The coefficient of $\mathbb{U}$ is 
\begin{equation}
	\frac{\mathbb{U}^{\prime }\mathbb{\bar{U}}^{\prime }-2a}{r}=
	\frac{2a}{m}\left(\frac{v^2}{2} - \frac{m}{r}\right)=\frac{2ah}{m},  
	\label{Keplerenergy}
\end{equation}
where $h$ is a constant of motion. Let us fix the parameter $a$ 
\begin{equation}
	a = -\frac{m}{2h},
\end{equation}
which means that $a$ is the semimajor axis of the elliptical motion. With this choice
the equation of motion is the harmonic oscillator with \textit{constant} frequency
\begin{equation}
	\mathbb{U}_{0}^{\prime \prime }+\mathbb{U}_{0}=0,
	\label{unpertubed equation}
\end{equation}%
where the subscript $0$ indicates the unperturbed case. 

The general solution
of this second order differential quaternion equation has six free
parameters as it is constrained with (\ref{KScondition}) and has a redundant
phase (\ref{gauge transformation}). The trial solution of the unperturbed 
case can be parametrized as follows 
\begin{equation}
	\mathbb{U}_{0}=a^{1/2}\beta_{+}^{-1/2}\left( q + \beta q^{-1}\right)
	\mathbb{R},  \label{unperturbed solution}
\end{equation}
where $\beta_{\pm}=1\pm\beta^{2}$ and $q=e^{\frac{\mathfrak{i}\omega}{2}}$ with $\omega =2\left( s-s_{p}\right)$. 
In the next paragraph it is shown that the trial solution describes an elliptical Kepler orbit.

Let us set $\mathbb{R}=1$ for the moment and substitute $\mathbb{U}_0$ into the 
definition of the separation
(\ref{transformation of the location})
\begin{equation}
	\mathbf{r}_0 =a \beta_{+}^{-1}\mathfrak{k}\left( z + \beta^2 z^{-1}+2 \beta\right)\,,  \label{aftersubstitution}
\end{equation}
where $z=q^{2}$ and the identity $q\mathfrak{k}=\mathfrak{k}q^{-1}$ was applied. This formula can be further 
expanded using the $z=\cos\omega+\mathfrak{i}\sin\omega$ identity
\begin{equation}
	\mathbf{r}_0 =a \mathfrak{k}(\cos\omega + 2 \beta \beta_{+}^{-1}) + a \mathfrak{j}\beta_{-}\beta_{+}^{-1}\sin\omega\,,  \label{rwithu}
\end{equation}
which means that according to (\ref{unitsaxis}) $\mathbf{r}_0$ describes the following parametric curve
\begin{align}
	x_0 &= a(\cos\omega + 2 \beta \beta_{+}^{-1})=a(\cos\omega+e)\,, \notag \\
	y_0 &= a\beta_{-}\beta_{+}^{-1}\sin\omega=a\sqrt{1-e^2}\sin\omega \,,
	\label{Kepler_orbit}
\end{align}
with $e=2\beta\beta_{+}^{-1}$. These equations describe an ellipse in the $(x,y)$ coordinate-plane,
with semimajor axis $a$, and eccentricity $e$. From equations (\ref{Kepler_orbit}) it follows that $\omega=0$
parametrizes the apocenter, thus $\omega$ is equivalent with the eccentric anomaly, except that the
latter is zero at the pericenter. This tells us that $s_p$ is the time advance of apocenter passage 
measured in modified time. In the general case $\mathbb{R}$ is obviously the rotation between the
orbital and reference frames, where the rotation according to (\ref{rotationsingeneral}) can be given with 
Euler angles 
\begin{equation} 
    \mathbb{R}=e^{\mathfrak{i}\frac{\psi}{2}}e^{\mathfrak{k}\frac{\theta}{2}}e^{\mathfrak{i}\frac{\phi}{2}}.
    \label{rotations}
\end{equation}

The formulae which provide the connection between the $a$, $\beta$, $s_p$ and angular parameters -- the 
orbital elements -- and the quaternion components are collected in the Appendix. 

Despite of the great advantages of the gauge condition (\ref{KScondition})
- which is especially fine for numerical calculations - for perturbative
calculations another geometrical condition is proposed.

\subsection{The perturbed case}

The trial solution for the perturbed equation of motion (\ref{regularized equation of motion}),
is just the unperturbed solution form (\ref{unperturbed
solution}) completed with general $\varepsilon$ proportional terms \cite{Vrbikbook}
\begin{equation}
\mathbb{U}=a^{1/2}\beta
_{+}^{-1/2}\left( q+\beta q^{-1}+q\mathbb{D}+\mathfrak{k}q\frac{\mathfrak{i}%
b+\mathbb{S}}{1+\beta z}\right)\mathbb{R} \,,
\label{the trial solution in the perturbed case}
\end{equation}%
where both the $\mathbb{D}$ and $\mathbb{S}$ quaternions are
$\mathcal{O}(\varepsilon )$ quantities, and complex in
the sense that they are in the subspace spanned by the units 1 and $
\mathfrak{i}$, while the $b$ quantity is real. 

Using the definition of the 
separation~(\ref{transformation of the location})
the perturbed separation vector in the Kepler frame is
\begin{equation}
        \mathbf{r}_K=\mathbf{r}_{K,0} + a \beta_{+}^{-1}\left\{ 
                   2\mathfrak{k}(z+\beta)\mathbb{D}
		-2\mathfrak{i}\,\text{Im}\,\mathbb{S}
		-2\mathfrak{i}\,b
		\right\}\,,  
	\label{pertaftersubstitution}
\end{equation}
where the $\varepsilon^2$ terms were neglected. The result tells us that parameter $b$
describes a translation along the $\mathfrak{i}$ unit which is a translation along the $z$-axis of
 the Kepler frame according to~(\ref{unitsaxis}). By considering~(\ref{rwithu}) it parametrizes a 
translation perpendicular to the orbital plane. The same is true for the imaginary part of $\mathbb{S}$, while 
the real part of $\mathbb{S}$ has no physical effect. Parameter $\mathbb{D}$ is complex and it is multiplied 
with $\mathfrak{k}$, thus the result is in the subspace spanned by the units $\mathfrak{j}$ and 
$\mathfrak{k}$. These units are associated~(\ref{unitsaxis}) with the $(x,y)$ coordinate plane of the 
Kepler frame, which is the orbital plane according to~(\ref{rwithu}). 

After introducing the trial solution in the perturbed case we fix the gauge. Vrbik`s
condition is that the real part of $\mathbb{S}$ must vanish\cite{Vrbikbook}
\begin{equation}
\mathbb{S^{\ast }}=-\mathbb{S}\,,  \label{gauge fixing in the perturbed case}
\end{equation}%
where the operator $^{\ast }$ conjugates its complex quaternion argument.
In this case the trial solution (\ref{the trial solution in the perturbed
case}) has no $\mathfrak{k}$ proportional part. 

Transformation (\ref{gauge
transformation}) has a simple geometrical interpretation. It describes a double 
rotation, one in the (1,$\mathfrak{k}$) and another one in the ($\mathfrak{i}$,$\mathfrak{j}$) 
subspace. Hence a transformation (\ref{gauge
transformation}) on $\mathbb{U}$ (\ref{the trial solution in the perturbed case}) whose tangent is the coefficient of the $\mathfrak{k}$ part of the
trial solution (\ref{the trial solution in the perturbed case}) divided by
its real part cancels the coefficient of $\mathfrak{k}$. In the leading $%
\varepsilon $ order this rotation is
\begin{equation}
\alpha =-\frac{\mathbb{S^{\ast }}+\mathbb{S}}{2(1+\beta z^{-1})(1+\beta z)}%
\,,  \label{gauge transform in the perturbed case}
\end{equation}%
which can be fulfilled without solving any differential equation for $\alpha
$ in contrary to (\ref{gamma trafo}).

\subsection{Example for the regularization}
\label{examplefortheregularization}
The one-dimensional two-body problem is considered with the following special
force $f=\varepsilon \dot{x}^{2}$, therefore
\begin{equation}
\ddot{x}+\frac{m }{x^{2}}=\varepsilon \dot{x}^{2}.  \label{kepler}
\end{equation}
We have chosen this kind of special force since the equation of motion has a constant of motion\footnote{
The first integral of the Eq. (\ref{kepler}) is $y^{\prime
2}=Cy^{2}e^{2\varepsilon y^{2}}/4+m /2+\varepsilon m
y^{2}e^{2\varepsilon y^{2}}Ei(-2\varepsilon y^{2})$,
where $C$ is the arbitrary constant of motion ($C=2E_{0}$ for unperturbed
motion) and $Ei(x)=-\underset{-x}{\overset{\infty }{\int }}$ $t^{-1}e^{-t}dt$
is the exponential integral function.}.

Eq. (\ref{kepler}) can be regularized with the following transformations
\begin{eqnarray}
x &=&y^{2},  \label{t1} \\
\frac{dt}{ds} &=&x.  \label{t2}
\end{eqnarray}
Then Eq. (\ref{kepler}) is%
\begin{equation}
y^{\prime \prime }+\frac{m -2(y^{\prime })^{2}}{2y}=2\varepsilon
y(y^{\prime })^{2}.  \label{pert}
\end{equation}%
In case of unperturbed motion ($\varepsilon =0$) the energy is the constant
of motion ($E_{0}=2(y^{\prime })^{2}/y^{2}-m /y^{2}$) and Eq. (\ref{pert})
is the equation of the harmonic oscillator

\begin{equation}
y_{0}^{\prime \prime }-\frac{E_{0}}{2}y_{0}=0\,,  \label{nullad}
\end{equation}%
where clearly for bounded motion $E_{0}<0$. The general solution for Eq. (\ref%
{nullad}) is $y_{0}=C_{1}e^{i\Omega s}+C_{2}e^{-i\Omega s}$, where $\Omega =%
\sqrt{\left\vert E_{0}\right\vert /2}$ is the orbital frequency.

\begin{figure}[tbh]
\begin{center}
\includegraphics[width=0.45\textwidth]{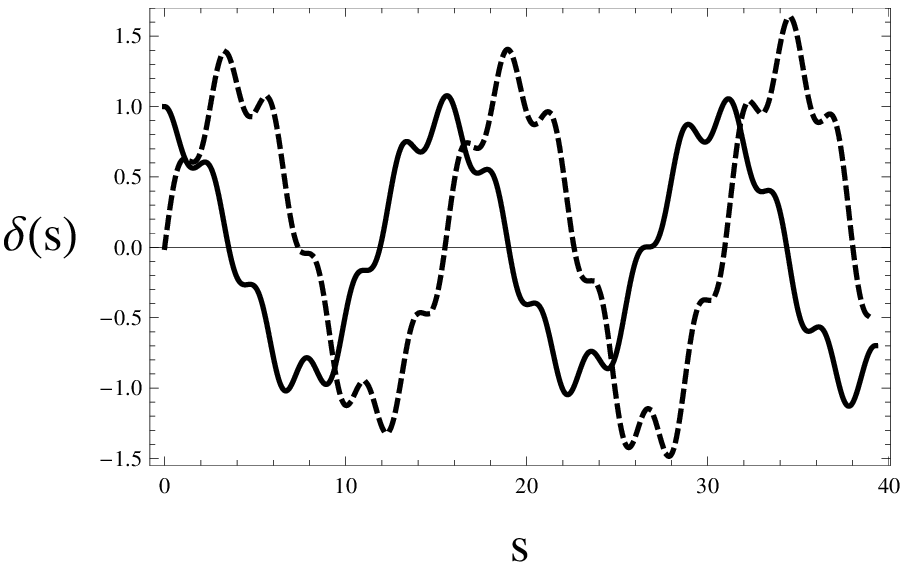} %
\includegraphics[width=0.45\textwidth]{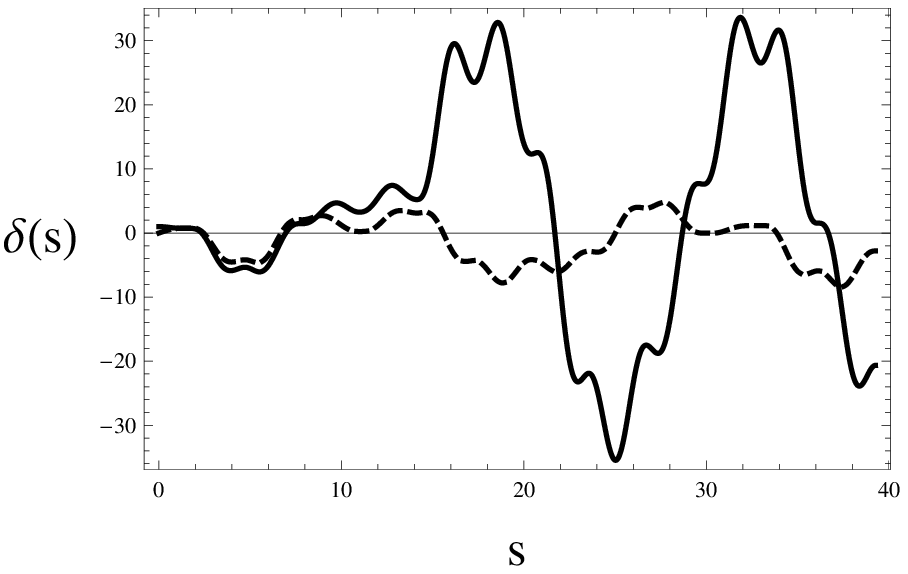}
\end{center}
\caption{The homogeneous (left) and inhomogeneous (right) solutions for Eq. (%
\protect\ref{elso2})$\ $for $A=1=\protect\Omega $, $\protect\delta ^{\prime
}(0)=0,\protect\delta (0)=1$ (dashed line) or $\protect\delta ^{\prime
}(0)=1,\protect\delta (0)=0$ (line).}
\label{hom}
\end{figure}

We assume that for perturbed motion ($\varepsilon \neq 0$) the form of the
solution is $y_{\varepsilon }=y_{0}+\varepsilon \delta $. Then Eq. (\ref%
{pert}) in the leading order of $\delta$ is
\begin{equation}
\delta ^{\prime \prime }-\frac{2y_{0}^{\prime }}{y_{0}}\delta ^{\prime }-%
\frac{E_{0}}{2}\delta =2y_{0}(y_{0}^{\prime })^{2},  \label{elso}
\end{equation}%
and using the $y_{0}=$ $A\cos (\Omega s)$ solution of the unperturbed motion
in (\ref{elso}) we get
\begin{equation}
\delta ^{\prime \prime }+2\Omega \tan (\Omega s)\delta ^{\prime }+\Omega
^{2}\delta =2\Omega ^{2}A^{3}\cos (\Omega s)\sin ^{2}(\Omega s),
\label{elso2}
\end{equation}%
where the sign of $\Omega ^{2}\delta $ is positive, since $E_{0}<0$.
Numerical solutions for Eq. (\ref{elso2}) can be seen on Fig. \ref{hom}. It
can be seen that the numerical solution $\delta(s)$ of these two examples are 
well-behaving, bounded functions for various initial values.

So far we have represented the regularization of the one-dimensional perturbed
two-body problem with a \textit{heuristic} special force. Let us consider the 
one-dimensional model using the \textit{generalized} 
Sundman transformation
\begin{equation}
	\frac{dt}{ds}=\tilde{x}\mathbf{,}  \label{t2b}
\end{equation}%
where $\tilde{x}$ is not fixed yet. Eq. (\ref{kepler}) can be regularized
using transformations (\ref{t1}) and (\ref{t2b})%
\begin{equation}
	y^{\prime \prime }+\frac{\left( y^{\prime }\right) ^{2}}{y}-\frac{\tilde{x}%
	^{\prime }y^{\prime }}{\tilde{x}}+\frac{m \tilde{x}^{2}}{2y^{5}}%
	=2\varepsilon y\left( y^{\prime }\right) ^{2}\mathbf{.}  \label{pp}
\end{equation}%
If $\tilde{x}=x$ (\textit{desingularized }in perturbed orbit) we obtain Eq. (%
\ref{elso}). If $\tilde{x}=x_{0}$ (\textit{desingularized }in unperturbed
orbit) the result is%
\begin{equation}
y^{\prime \prime }+\frac{\left( y^{\prime }\right) ^{2}}{y}-\frac{%
x_{0}^{\prime }y^{\prime }}{x_{0}}+\frac{m x_{0}^{2}}{2y^{5}}=2\varepsilon
y\left( y^{\prime }\right) ^{2}\mathbf{.}  \label{perti}
\end{equation}%
Substituting the $y_{\varepsilon }=y_{0}+\varepsilon \delta _{0}$ perturbed
solution, one obtains
\begin{equation}
\delta _{0}^{\prime \prime }-\frac{E(y_{0})}{2}\delta _{0}=2\varepsilon
y_{0}\left( y_{0}^{\prime }\right) ^{2},  \label{uj}
\end{equation}%
where $E(y_{0})=[2\left( y_{0}^{\prime }\right) ^{2}+5 m ]/y_{0}^{2}$ is
not the constant of motion (note that the coefficient of the linear term is
the constant of motion in case $\tilde{x}=x$ (Eq. (\ref{elso})).

\begin{figure}[tbh]
\begin{center}
\includegraphics[width=0.43\textwidth]{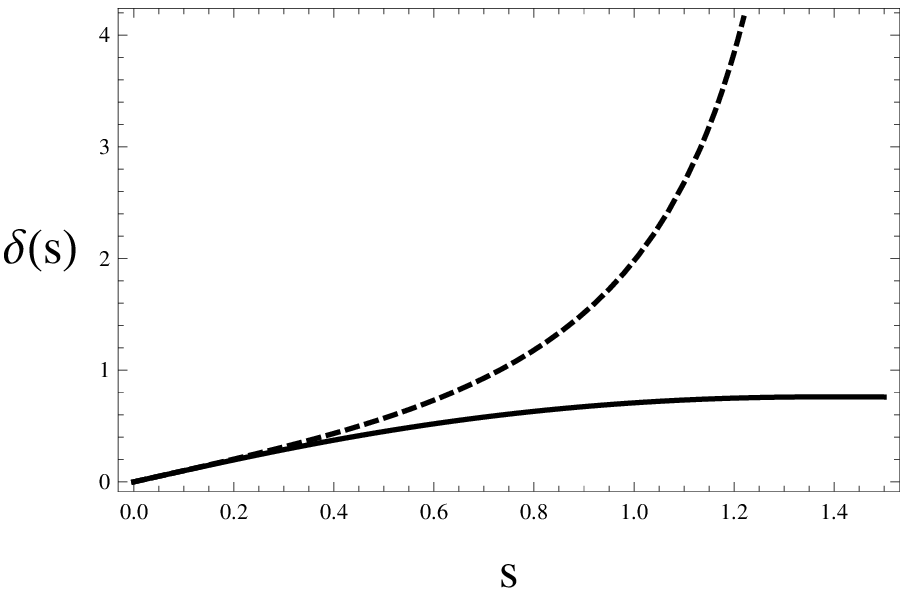} 
\includegraphics[width=0.43\textwidth]{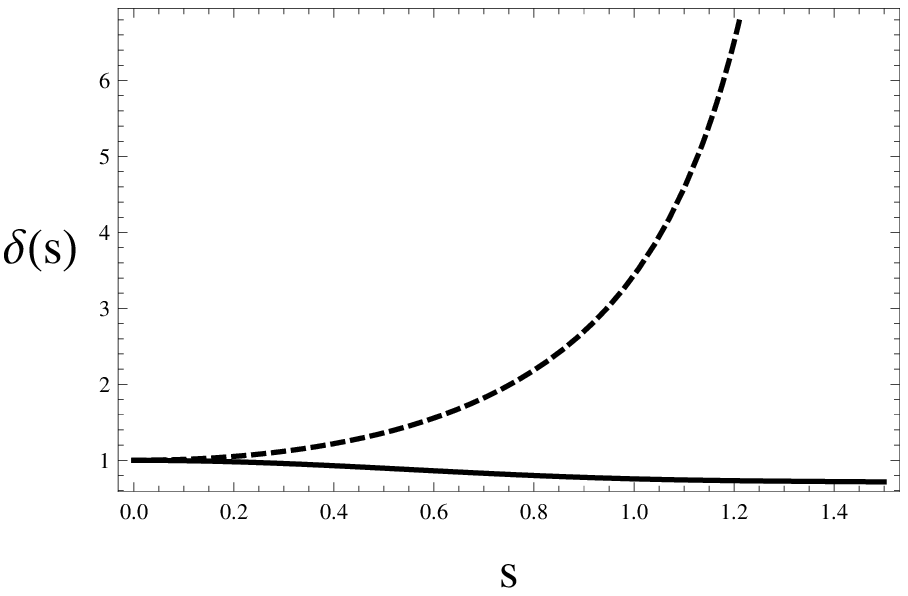}
\label{bifur}
\end{center}
\caption{The numerical solutions for (\protect\ref{a1}) and (\protect\ref{a2}%
) with different initial values}
\end{figure}

In total two types of desingularization ($\tilde{x}=x,x_{0}$) using the $
y_{0}=$ $A\cos (\Omega s)$ (we have fixed the frequency $\Omega =1$)
unperturbed solution can be given
\begin{eqnarray}
\delta ^{\prime \prime }+2\tan (s)\delta ^{\prime }+\delta &=&2A^{3}\cos
s\sin ^{2}s,  \label{a1} \\
\delta _{0}^{\prime \prime }-\left( \frac{5 m }{2A^{2}}\sec ^{2}s+\tan
^{2}s\right) \delta _{0} &=&2A^{3}\cos s\sin ^{2}s.  \label{a2}
\end{eqnarray}
The numerical analyzis of these two equations with different initial
values can be seen on Fig. \ref{bifur}.

It can be seen that in this one-dimensional perturbed two-body problem the two
types of desingularization methods lead to quite different solutions. Therefore 
the Sundman transformation is generally nontrivial in perturbed equations.

\section{Orbital elements with quaternions}
\label{Orbitalelementswithquaternions}
Before explaining the quaternion approach the classical equations are
described in order to explain the relationship between the two different methods. 
The equation system of the classical two-body problem is of total order six, hence it can be described with six first
integrals, which are also called \textit{orbital element}s. These elements are the
\textit{semi-major axis} $a$, the \textit{eccentricity} $e$, the \textit{%
inclination }$\theta $, \textit{longitude of the ascending node} $\phi$, the
\textit{argument of the pericenter} $\psi $ \footnote{in classical celestial
mechanics the symbols are $\iota $, $\Omega $ and $\omega $ respectively. Here we adopted the notations of J.Vrbik.} and the%
\textit{\ mean anomaly at the epoch} $l_{0}$ (or time of pericenter passage $%
t_{0}$) related to~the dynamics. The \textit{Lagrange planetary equation}s
in the standard perturbed two-body problem are\cite{Brumberg}
\begin{eqnarray}
	\frac{da}{dt} &=&\frac{2}{n\sqrt{1-e^{2}}}\left( Se\sin \chi +T\frac{a\left(
	1-e^{2}\right) }{r}\right) ,  \nonumber \\
	\frac{de}{dt} &=&\frac{\sqrt{1-e^{2}}}{na}\left[ S\sin \chi +T\left( \cos
	\chi +\cos \xi \right) \right] ,  \nonumber \\
	\frac{d\theta }{dt} &=&\frac{r\cos (\chi +\psi )}{na^{2}\sqrt{1-e^{2}}}W, 
	\label{Lagrange_equations} \nonumber\\
	\frac{d\phi }{dt} &=&\frac{r\sin (\chi +\psi )}{na^{2}\sqrt{1-e^{2}}\sin
	\theta }W,  \nonumber \\
	\frac{d\psi }{dt} &=&-\cos \theta \frac{d\phi }{dt}+\frac{\sqrt{1-e^{2}}}{nae
	}\left[ T\left( 1+\frac{r}{a\left( 1-e^{2}\right) }\right) \sin \chi -S\cos
	\chi \right] ,  \nonumber \\
	\frac{dl_{0}}{dt} &=&-\sqrt{1-e^{2}}\left( \frac{d\psi }{dt}+\cos \theta 
	\frac{d\phi }{dt}\right) -S\frac{2r}{na^{2}},
\end{eqnarray}
where $\chi $ is the \textit{true anomaly}, $\xi $ is the \textit{eccentric
anomaly} and $r$ is the parametrization of the osculating orbit
\begin{equation}
	r=\frac{a(1-e^{2})}{1+e\cos \chi }=a(1-e\cos \xi ), 
\end{equation}
and $l$ is the \textit{mean anomaly}, which can be defined by the
Kepler equation%
\begin{equation}
	l-l_{0}=n\left( t-t_{0}\right) =\xi -e\sin \xi ,
\end{equation}%
and $n=m^{1/2}a^{-3/2}$ is the \textit{mean motion}. 

The $S$, $T$ quantities are the projections of the perturbing
force to the orbital plane, while $W$ is the projection to the normal vector
of the orbital plane $\mathbf{\hat{k}}$
\begin{equation}
S=\mathbf{\hat{r}\cdot f,}\qquad T=(\mathbf{\hat{k}}\times \mathbf{\hat{r}%
)\cdot f,}\qquad W=\mathbf{\hat{k}}\cdot \mathbf{f}\,.  \label{auxquantity}
\end{equation}

To derive \textit{quaternion} differential equations for the orbital elements the
trial solution (\ref{the trial solution in the perturbed case}) has to be
substituted into the equation of motion (\ref{regularized equation of motion}).
To simplify the calculation the quaternion equation of motion (\ref{regularized equation of motion}) 
is decoupled into two complex equations.
The derivation of the complex equations are given in details\cite{Vrbikbook,Notebooks} and the most important
steps are briefly outlined in our Appendix. Here only the solution and the necessary definitions are presented. 

The following auxiliary quaternion quantities have to be defined
\begin{align}
\mathbb{Q}&=-2\varepsilon \frac{a}{m}\frac{Cx\left( \mathbf{r}_{K}%
\mathbf{f}_{K}\right) }{1+\beta z}=-2\varepsilon \frac{a}{m }\frac{%
{}Cx\left( \mathbf{r}_{K0}\mathbf{f}_{K0}\right) }{1+\beta z}+O\left(
\varepsilon ^{2}\right), \notag \\ 
\mathbb{W}&=-4\varepsilon \frac{a}{m}r Cx\left( \mathbf{f}_{K}\right)
=-4\varepsilon \frac{a}{m}r_{0}Cx\left( \mathbf{f}_{K0}\right) +O\left(
\varepsilon ^{2}\right),    
\label{WQdef}
\end{align}
where the operator $Cx$ is a projector, which projects its quaternion argument to the complex subspace
spanned by the units $1$ and $\mathfrak{i}$. The additional subscript $0$
indicates the unperturbed value of the symbol. The subscript $K$ is omitted
in $r$ and $r_0$ as they are scalars and have the same value in every frame.
To point out the relationship between the quaternion formulae and the classical equations (\ref{Lagrange_equations}) note
that the real and imaginary part of the complex $\mathbb{Q}$ quantity is
proportional to the previously introduced $S$ and $T$ ($\ref{auxquantity}$) respectively, 
while $\mathbb{W}$ is proportional to $W$. 

The quaternion coefficients (\ref{WQdef}) can be expanded into Laurent series\cite{Vrbikbook} 
\begin{equation}
	\mathbb{Q}=\sum_{n=-\infty }^{n=+\infty }\mathcal{Q}_{n}z^{n}\,,\;\;
	\mathbb{W}=\sum_{n=-\infty }^{n=+\infty }\mathcal{W}_{n}z^{n}\,,
	\label{QW}
\end{equation}
together with $\mathbb{D}$ and $\mathbb{S}$ from (\ref{the trial solution in the perturbed case})
\begin{equation}
        \mathbb{D}=\sum_{\substack{n=-\infty \\ n \ne -1,0 }}^{n=+\infty }\mathcal{D}_{n}z^{n}\,,\;\;
        \mathbb{S}=\sum_{n=2 }^{n=+\infty }\mathcal{S}_{n}z^{n}\,.
        \label{DS}
\end{equation}
The Laurent series are given in powers of $z$. From
the definition of the orbital separation $\mathbf{r}$ (\ref{transformation of the location}) follows          
that this is enough as the expansion of the separation contains every power of $q$. The 
coefficients $\mathcal{D}_{-1}$, $\mathcal{D}_{0}$ and $\mathcal{S}_{-1}$, $\mathcal{S}_{1}$
were left out from the expansion of $\mathbb{D}$ and $\mathbb{S}$ since they would 
only duplicate the $q$ and $q^{-1}$ terms of the solution (\ref{the trial solution in the perturbed case}),
while $\mathcal{S}_0$ was explicitly separated as $b$.\par
Substituting expansions (\ref{QW}) and (\ref{DS}) into the complex equations (\ref%
{mozgegyenlet 1i 2}) and (\ref{mozg jk}) the differential equations for the
orbital elements can be extracted by matching the coefficients of $z$ with the same power on 
both side of the equation. The obtained differential equations are the following\cite{Vrbikbook,Notebooks}
\begin{eqnarray}
a^{\prime } &=&2a\,\text{Im}\left( \mathcal{Q}_{0}-\beta \mathcal{Q}_{-1}\right) ,
\label{eredmenyek} \\
\beta ^{\prime } &=&-\frac{\beta _{+}}{4}\text{Im}\left( \mathcal{Q}_{1}+3\beta
\mathcal{Q}_{0}+3\mathcal{Q}_{-1}+\beta \mathcal{Q}_{-2}\right) ,
\label{dbe} \\
Z_{1} &=&-\beta _{-}^{-1}\text{Im}\left( \frac{\beta _{+}}{2}\mathcal{W}_{1}+\beta
\mathcal{W}_{0}\right) ,  \label{z1e} \\
Z_{2} &=&-\frac{1}{2}\text{Re}\left( \mathcal{W}_{1}\right) ,  \label{z2e} \\
Z_{3} &=&\frac{1}{4\beta }\text{Re}\left\{ -\beta _{+}\mathcal{Q}_{1}+\beta \left(
1-3\beta ^{2}\right) \mathcal{Q}_{0}+\left( 3-\beta ^{2}\right) \mathcal{Q}%
_{-1}\right.  \notag \\
&&\left. +\beta \beta _{+}\mathcal{Q}_{-2}\right\},  \label{z3e} \\
s_{p}^{\prime } &=&\frac{Z_{3}}{2}+\frac{\beta _{+}^{-1}}{4}\text{Re}\left\{ \beta
\left( 2+\beta ^{2}\right) \mathcal{Q}_{1}+\left( \beta _{+}+3\beta
^{4}\right) \mathcal{Q}_{0}\right.  \notag \\
&&\left. -\beta \left( 1-2\beta ^{2}\right) \mathcal{Q}_{-1}-\beta ^{4}%
\mathcal{Q}_{-2}\right\} ,  \label{eddig} \\
b &=&\frac{1}{8}\text{Im}\left\{ \left( \beta _{-}^{2}\mathcal{W}_{0}+2\beta ^{2}%
\mathcal{W}_{2}\right) \beta _{+}^{-1}+\beta \mathcal{W}_{1}\right\} ,
\label{Es megeddig}
\end{eqnarray}%
and the two additional formulae for $\mathbb{D}$ and $\mathbb{S}$ is given in
the Appendix. The $Z_i$ quantities are the components of $\mathbb{Z}_K$, where
the Kepler frame subscript was dropped to simplify the notation. Note that
equation (\ref{z3e}) is singular in $\beta $, which shows that in the
circular orbit limit the ordinary sense of the rotation no longer valid.

The coefficients $\mathcal{Q}_n$ or $\mathcal{W}_n$ of the Laurent series can be obtained with a contour
integral, where $C_0$ is the unit circle 
\begin{equation}
	\mathcal{Q}_n=\oint_{C_0}\frac{\mathbb{Q}}{z^n}\frac{dz}{2\pi\mathfrak{i}\,z}\,.
	\label{qnint}
\end{equation}

Note that the Laurent expansion (\ref{DS}) of $\mathbb{D}$ and $\mathbb{S}$ has simplified
the form of the differential equations (\ref{eredmenyek})-(\ref{Es megeddig}) with respect to the 
Lagrange's planetary equations (\ref{Lagrange_equations}). The Laurent series of $\mathbb{D}$ and $\mathbb{S}$
absorbed the ``short" term oscillatory part of the equation. The remaining differential equations contain only
the adiabatic, ``long" term part, which might be easier to solve.

We have to amend the equations above with the transformation of the angular
velocity from the comoving Kepler frame to the inertial system, which are
the following
\begin{eqnarray}
\phi ^{\prime } &=&\frac{Z_{1}\sin \psi +Z_{2}\cos \psi }{\sin \theta }%
\mathbf{,\;\;}  \notag \\
\theta ^{\prime } &=&Z_{1}\cos \psi -Z_{2}\sin \psi ,  \notag \\
\psi ^{\prime } &=&Z_{3}-\phi ^{\prime }\cos \theta .  \label{kepler-inertia}
\end{eqnarray}%
This transformation is familiar from classical mechanics, in deriving the
Euler equations of the the rigid body.

\section{GR perturbations}
\label{GRperturbations}
In this section perturbations calculated from the general relativity 
are examined using the described quaternion approach. The perturbations are
examined with post-Newtonian approximation. The post-Newtonian approximation applies an expansion of
corrections to the Newtonian gravitational theory with an expansion parameter
$\varepsilon \approx v^2 \approx m/r$, which is supposed to be small, where $v$ is the 
velocity.

We use equations up to $\varepsilon^{5/2}$, (post)$^{5/2}$-Newtonian 
order, which is the order where the dominant gravitational radiation damping forces occur.   
   
First of all, the (post)$^1$-Newtonian correction to the classical mechanics will be examined in the first section. This
is followed by the (post)$^{5/2}$-Newtonian analysis of gravitational radiation where we rederive the classical Peters-Mathews formula.  

\subsection{Planar assumption}
The mentioned perturbations are \textit{planar} perturbations, in the sense that the force lies
within the orbital plane. In this case obviously $\mathbb{S}=b=0$ and the trial solution (\ref{the trial solution in the perturbed case})
contains only
perturbations within the orbital plane
\begin{equation}
	\mathbb{U}_{K}=a^{1/2}\beta_{+}^{-1/2}\left( q+\beta q^{-1}+q\mathbb{D}\right)\,,
	\label{the_trial_solution_planar_case}
\end{equation}

It follows that in case of planar forces the $\mathbb{S}^*=-\mathbb{S}$ condition 
(\ref{gauge fixing in the perturbed case}) is true. Remarkably  
the $\Gamma=0$ condition is also satisfied\cite{Vrbikbook}. To show this we need the derivative of $\mathbb{U}$
expressed with Kepler frame quantities 
\begin{equation}
	\mathbb{U}'=\mathbb{U}_K'\mathbb{R}+\mathbb{U}_K\mathbb{R}'=
                    \left(\mathbb{U}_K'+ \mathbb{U}_K \frac{\mathbb{Z}_K}{2}\right)\mathbb{R}.
\end{equation}
Therefore
\begin{equation}
	\Gamma=2\,\text{Re} \left(\bar{\mathbb{U}}\mathfrak{k}\mathbb{U}'\right)=
		2\,\text{Re} \left\{\bar{\mathbb{R}}\bar{\mathbb{U}}_K \mathfrak{k} 
		\left(\mathbb{U}_K'+ \mathbb{U}_K \frac{\mathbb{Z}_K}{2}\right)\mathbb{R}\right\}\,,
\end{equation}
and since the rotation can be dropped under the real part operator
\begin{equation}
        \Gamma = 2\,\text{Re} \left(\bar{\mathbb{U}}_K \mathfrak{k} \mathbb{U}_K'+ 
        \mathbf{r}_K \frac{\mathbb{Z}_K}{2}\right)\,.
	\label{gammainkepler}
\end{equation}

In the planar case $\mathbb{U}_K$ is a complex number, therefore both $\bar{\mathbb{U}}_K \mathfrak{k} \mathbb{U}_K'$ and $\mathbf{r}_K$ are 
in the orbital plane spanned by the $\mathfrak{j}$ and $\mathfrak{k}$ units. In the planar case 
the orbital plane is preserved,  therefore $\mathbb{Z}_K$ must be perpendicular to this $\mathfrak{j}$, $\mathfrak{k}$ subspace. It means 
that $\mathbb{Z}_K$ has only $\mathfrak{i}$ part. It is easy to see from equation (\ref{gammainkepler}) that the argument of the
operator $\text{Re}$ has no real part. Therefore in case of planar forces $\Gamma$ vanishes. 

Consequently the equation of motion (\ref{regularized equation of motion}) simplifies
\begin{equation}
	2\mathbb{U}^{\prime \prime }-\left( 2\mathbb{U}^{\prime }\overline{\mathbb{U}%
	^{\prime }}-4a\right) \frac{\mathbb{U}}{r}
	-\mathbb{U}^{\prime }\frac{a^{\prime }}{a}+4\frac{a}{m }\varepsilon \mathbb{U}%
	\mathbf{rf}=0\,.
\end{equation}
Let us introduce a $\tau$ parameter by rescaling the modified time $d\tau=2a^{1/2}m^{-1/2}ds$. With the
help of the $\tau$ parameter the equation of motion is just the perturbed harmonic oscillator
\begin{equation}
	2\frac{d^2\mathbb{U}}{d\tau^2} - h \mathbb{U} + \varepsilon \mathbb{U}\mathbf{rf}=0\,.
\end{equation}
In the planar case the equation of motion 
substantially simplified and identical with the equation of Waldvogel\cite{Waldvogel}.\footnote{Only the sign convention
of $h$ is different.}

In the planar case the perturbations $\mathbb{D}$ can be expressed in a more conventional
way. Let us introduce $a=a_0+\delta a$ and $\beta=\beta_0+\delta\beta$ in
the trial solution (\ref{unperturbed solution}) where $\delta a$ and $\delta\beta$ are first order quantities.
In this case by matching the first order part of Eqs. (\ref{unperturbed solution}) and (\ref{the_trial_solution_planar_case})
one obtaines the following important relations
\begin{align}
	\delta a=2 a_0 \frac{\text{Im} \left(  B\,\mathbb{D} \right)}{\text{Im}\left(A^*B\right)}\,,\,   
	\delta\beta=2\frac{\text{Im}\left(A\,\mathbb{D}\right)}{z^{-1}-z}\,,
\end{align}
where $A=1+\beta_0 z$ and $B=z-\beta_0(1-\beta_0)^2 A$.

\subsection{The classical post-Newtonian effect}
In this section the leading contribution of general relativity to classical Newtonian mechanics
is examined in details. The force is given by numerous authors \cite{Kidder:1995zr}
\begin{equation}
	\mathbf{a}_{PN}=-\frac{m}{r^{2}}\left\{ \mathbf{\hat{n}}\left[ \left(
	1+3\eta \right) v^{2}-2\left( 2+\eta \right) \frac{m}{r}-\frac{3}{2}\eta
	\dot{r}^{2}\right] -2\left( 2-\eta \right) \dot{r}\mathbf{v}\right\} .
	\label{post newtonian effect}
\end{equation}%
where the subscript $PN$ denotes the post-Newtonian term, $\mathbf{\hat{n}}=\mathbf{r/}r$, $\eta=(m_1 m_2)/m^2$ 
and $v=\left\vert \mathbf{v}\right\vert $ is 
the absolute value of the orbital velocity $\mathbf{v=}d\mathbf{r/}dt$.

After transforming it to quaternion expression 
\begin{align}
	\mathbf{a}_{PN}=&-\frac{\mathbf{r}_{K}}{r^{5}}\frac{m}{4a}(1+3\eta )\left(
	\mathbf{r}_{K}^{\prime }\mathbf{r}_{K}^{\prime }\right) +\frac{\mathbf{r}_{K}
	}{r^{4}}2(2+\eta )m^{2}  \label{pn1 quaternio} \nonumber\\
	&+\frac{\mathbf{r}_{K}}{r^{5}}\frac{3}{8}\frac{m^{2}\eta }{a}\left(
	r^{\prime }\right) ^{2}+\frac{\mathbf{r}_{K}^{\prime }}{r^{4}}\frac{m^{2}}{2a
	}(2-\eta )r^{\prime }. 
\end{align}
This result (\ref{pn1 quaternio}) have to be substituted into the definition of $\mathbb{Q}$ (\ref{WQdef}), where 
the separation $\mathbf{r}_K$, its magnitude $r$ and their derivatives are treated as functions of $z$ according to 
(\ref{aftersubstitution}). Therefore the result is a function of $z$
\begin{eqnarray}
	\mathbb{Q} &=&\frac{mz\beta _{+}}{a(z+\beta )^{3}(1+z\beta )^{4}}\left[
	8z^{3}\beta \left( 2\beta ^{2}+\eta \right) +\beta ^{2}(1+z^{4})(7\eta-6)\right.
	\label{pn1 Q} \nonumber\\
	&&\left. +8z\beta \left( 2+\beta ^{2}\eta \right)
	+z^{2}\left\{ 6-2\beta ^{4}(\eta-3)-2\eta +\beta ^{2}(32+6\eta )\right\}
	\right]\,.
\end{eqnarray}

Applying the contour integral (\ref{qnint}) the coefficients $\mathcal{Q}_n$ can be computed as follows. $\beta<1$ therefore
the only singularity of $\mathbb{Q}$ inside $C_0$ is at $-\beta$. The other pole at $-1/\beta$ lies outside the 
unit circle.

$\mathbb{Q}$ can be expanded around its pole at $-\beta$ and
keeping the coefficient of the $(z+\beta)^{-1}$ part, the result is
\begin{equation}
	\mathcal{Q}_{-1}=2ma^{-1}\beta _{-}^{-4}\beta _{+}\left( -\beta -8\beta
	^{3}-3\beta ^{5}+3\beta \eta +17\beta ^{3}\eta +\beta ^{5}\eta \right)\,.
	\label{qm1}
\end{equation}
In the same way with $\mathbb{Q}/z$ one finds
\begin{equation}
	Q_{0}=-2ma^{-1}\beta _{-}^{-4}\beta _{+}\left( -3-8\beta ^{2}-\beta
	^{4}+\eta +17\beta ^{2}\eta +3\beta ^{4}\eta \right)\,.
	\label{q0}
\end{equation}
Both of the coefficients $\mathcal{Q}_{-1}$ and $\mathcal{Q}_0$ are real. The differential equation for 
the semimajor axis is proportional to their imaginary part (\ref{eredmenyek}), therefore $a'=0$. The remaining 
differential equations can be calculated in the same way.

The resulting nontrivial differential equations in modified time for the
orbital parameters are as follows\cite{Notebooks}. The equation for the argument of the
pericenter
\begin{equation}
        \psi ^{\prime }=\frac{6m}{a}\left(\frac{\beta_{+}}{\beta_{-}}\right)^2,  \label{post newtonian effect psi}
\end{equation}
and for the modified time at apocenter
\begin{equation}
	s_{p}^{\prime }=-\frac{m}{2 a \beta _{-}}\left( \eta -9+\beta ^{2}\left( 8\eta
	-15\right) \right).  \label{post newtonian effect sp}
\end{equation}%
Using the transformation to the modified time (\ref{transformation of the
time}) in the leading order
\begin{equation*}
	\frac{d}{dt}=\frac{1}{2a}\sqrt{\frac{m}{a}}\frac{d}{ds}\,,
\end{equation*}%
from (\ref{post newtonian effect psi}) it follows that
\begin{equation}
	\dot{\psi}=\frac{3m^{3/2}}{a^{5/2}\left( 1-e^{2}\right) },  \label{merkur}
\end{equation}%
which is the known expression for the precession of the pericenter\cite{Landau II}. The
remaining differential equations have zero on the right hand side of the equation and
the corresponding orbital element is constant.

\subsection{Gravitational radiation reaction}

\label{GR radiation} 
Gravitational radiation damping has been recognized as a process with very important 
observable consequences: the PSR 1913+16 system has given the first evidence that
gravitational waves exist\cite{Taylor:1991yt}, and other systems are of high
importance as well \cite{HT cikk,Kramer:2005ez}. The equation of motion is given by \cite{Iyer}
\begin{equation}
\mathbf{a}_{RR}=-\frac{8\eta m^2}{5r^3}\left(-A_{5/2} \dot{r}\mathbf{\hat{n}}
+ B_{5/2}\mathbf{v} \right) ,  \label{RR a rr}
\end{equation}
where the subscript $RR$ indicates the radiation reaction term and
\footnote{
Our notation is slightly different from \cite{Iyer} as the $\alpha$ and $
\beta$ parameters are occupied; instead we use $\gamma$ and $\rho$
respectively.}
\begin{align}  \label{IWtrafo}
A_{5/2} &= 3(1+\rho)v^2+\frac{1}{3}(23 + 6\gamma-9\rho)\frac{m}{r} - 5\rho
\dot{r}^2 \nonumber\\
B_{5/2} &= (2+\gamma)v^2+(2-\gamma)\frac{m}{r}-3(1+\gamma)\dot{r}^2.  
\end{align}

The $\gamma $ and $\rho $ parameters in $\left( \ref{IWtrafo}\right) $
represent the residue of gauge freedom that has not been fixed by the energy
balance method and that has no physical meaning. It is known that these
arbitrariness is equivalent with a coordinate transformation whose effect on
the two-body separation vector is
\begin{equation}
\mathbf{r}\rightarrow \mathbf{r}+\mathbf{\delta r}=\mathbf{r}+ \frac{8 \eta
m^2}{15 r^2}[\rho \dot{r}\mathbf{r}+\left( 2\rho -3\gamma \right) r\mathbf{{v%
}].}  \label{gaugedep_with_r}
\end{equation}%
We use transformation (\ref{gaugedep_with_r}) to \textit{remove} the gauge
dependency from the quaternion equation (\ref{regularized equation of motion}%
), after substituting (\ref{RR a rr}) as the perturbing force.

In order to apply transformation (\ref{gaugedep_with_r}) on the quaternion equation
of motion (\ref{regularized equation of motion}) we have to rewrite it in 
quaternion form using modified time (\ref{transformation of the time}). The definition of the
modified time (\ref{transformation of the time}) contains the separation $r$
therefore any gauge dependent transformation of the separation, like (\ref%
{gaugedep_with_r}), leads to gauge dependent modified time $s\left( \gamma
,\rho \right) $. Consequently the transformation of any real time derivative
involves a new gauge dependent contribution
\begin{equation}
\frac{d}{dt}=\sqrt{\frac{m }{a}}\frac{1}{2r}\frac{d}{ds}\rightarrow \sqrt{%
\frac{m}{a}}\frac{1}{2r}\frac{d}{ds}-\sqrt{\frac{m}{a}}\frac{1}{4r^{2}}%
\delta r\frac{d}{ds}+{\mathcal{O}\left( \delta r^{2}\right) }.
\label{transformation of the time2}
\end{equation}%
It follows that transformation (\ref{gaugedep_with_r}) in its original form
does not cancel these new gauge dependent contributions and needs to be
reparametrized. The reparametrized transformation in quaternion form is
the following
\begin{equation}
{\mathbb{U}}_{K}\rightarrow {\mathbb{U}}_{K}+\frac{2\eta m^{5/2}}{%
15r^{3}a^{1/2}}[K\,\rho \,r^{\prime }{\mathbb{U}}_{K}+\left( L\,\rho
-N\,\gamma \right) r\,{\mathbb{U}}_{K}^{\prime }],  \label{gaugedep_with_U}
\end{equation}%
where $K$, $L$ and $N$ are unknown coefficients. 

To obtain them consider that
$\mathbb{Q}$ (\ref{WQdef}) is Laurent series in $z$ and gauge independence
requires that every $\rho $ and $\gamma $ proportional term in the
coefficients must vanish. E.g. the coefficients of $z^{2}\gamma $, $%
z^{2}\rho $ and $z^{3}\beta ^{2}\rho $ of $\mathbb{Q}$ (\ref{WQdef}) after
simplification lead to a linear equation system which determines that $N=3$ 
and $K=L=1$ \cite{Notebooks}. In the leading order according 
to (\ref{transformation of the location}) this is equivalent with the following 
real time vectorial transformation
\begin{equation}
\mathbf{r}\rightarrow \mathbf{r}+\frac{8\eta m^{2}}{15r^{2}}[\rho \dot{r}%
\mathbf{r}+\frac{1}{2}\left( \rho -3\gamma \right) r\mathbf{{v}],}
\label{gaugedep_with_r2}
\end{equation}%
which is slightly different from (\ref{gaugedep_with_r}).

The result from formula (\ref{eredmenyek})-(\ref{Es megeddig}) for the
semimajor axis is now gauge independent\cite{Notebooks}
\begin{equation}
a^{\prime }=-\frac{64m^{5/2}\eta \beta _{+}^{3}\beta _{-}^{-7}}{15a^{3/2}}%
\left( 6+97\beta ^{2}+219\beta ^{4}+97\beta ^{6}+6\beta ^{8}\right) ,
\label{gravitational radiation result a}
\end{equation}%
and also for the modified eccentricity
\begin{equation}
\beta ^{\prime }=-\frac{8m^{5/2}\eta\beta \beta _{+}^{4}\beta _{-}^{-6}}{%
15a^{5/2}}\left( 76+273\beta ^{2}+76\beta ^{4}\right) .
\label{gravitational radiation result beta}
\end{equation}%
The remaining differential equations are trivial, with a zero on the right
hand side, and the corresponding orbital elements remain constant. The gauge
independent value of parameter ${\mathbb{D}}$ is given in \cite{Notebooks},
while ${\mathbb{S}}$ is zero.

Now we are in the position that the latter result for the semi major axis (%
\ref{gravitational radiation result a}) and also the expression for the
eccentricity (\ref{gravitational radiation result beta}) can be easily
verified. They must be equal with the two corresponding classical formula
derived from the well known Peters-Mathews formula \cite{Peters:1963ux},
which describes the effect of gravitational radiation. After substituting the
expression $e=2\beta \beta _{+}^{-1}$ into (\ref{RR fel nagytengely}) and (%
\ref{RR excentricitas}) together with the transformation rule between the
real and modified time (\ref{transformation of the time}) one can derive the
two equation below
\begin{eqnarray}
\frac{da}{dt} &=&-\frac{64}{5}\frac{\eta m^{3}}{a^{3}}\frac{1}{(1-e^{2})^{7/2}}%
\left( 1+\frac{73}{24}e^{2}+\frac{37}{96}e^{4}\right) ,
\label{RR fel nagytengely} \\
\frac{de}{dt} &=&-\frac{304}{15}\frac{\eta m^{3}}{a^{4}}\frac{e}{\left(
1-e^{2}\right) ^{5/2}}\left( 1+\frac{121}{304}e^{2}\right) ,
\label{RR excentricitas}
\end{eqnarray}%
which are indeed identical with the two formula derived from the
Peters-Mathews equation \cite{Magiore}.

\section{Conclusion and outlook}

\label{Conclusion} 
In this paper general relativity perturbations were examined using a new approach
where the regularization of the Kepler problem is given with quaternions. This approach 
is based on the usual Kustaanheimo-Stiefel method which is defined with matrices. 

With the new calculus the differential equations of the orbital parameters were derived
in case when the perturbation is the leading (post)$^{1}$-Newtonian order correction of 
general relativity. To test the new method the precession of the pericentre is rederived. 

Then the gravitational radiation reaction was analyzed, where the famous Peters-Mathews formula was
reproved using the quaternion approach \cite{Peters:1963ux}. We have studied the gauge dependence of 
the equations of motion and we managed to remove the residual gauge freedom from the quaternionic equation 
of motion. 

The new quaternionic approach is easy to implement with program code. Quaternions can be represented 
with pairs of complex numbers, then the equations can be calculated and solved with the help of
complex analysis. This feature makes this method to a very efficient calculus for symbolic 
computations.  

With the quaternion based regularization the spin-orbit and spin-spin interactions can be examined
as well\cite{Barker:1975ae}. It is foreseen that these spin interaction related calculations provide the next step of 
our studies.

\section*{Acknowledgment}
We would like to thank Prof. J. Vrbik to explain and interpret some points 
of his method.
F. N. especially would like to thank Prof. P. Forg\'{a}cs for the initiation of 
this study and for his help during the whole work. The authors would like to
thank M. Vas\'{u}th for giving valuable information and advices.

\appendix
\section{The components of $\mathbb{U}$ expressed with orbital elements}
These formulae can be straightforwardly derived from the unperturbed
solution~(\ref{unperturbed solution}) using the expression of the rotation~(%
\ref{rotations}) with the rotation angles
\begin{align}
U& =\phantom{-}a^{1/2}\beta _{+}^{-1/2}\cos \frac{\theta }{2}\left\{ \cos \left( \omega
_{+}+\frac{\omega }{2}\right) +\beta \cos \left( \omega _{+}-\frac{\omega }{2%
}\right) \right\} \,,  \notag \\
U_{3}& =\phantom{-}a^{1/2}\beta _{+}^{-1/2}\cos \frac{\theta }{2}\left\{ \sin \left(
\omega _{+}+\frac{\omega }{2}\right) +\beta \sin \left( \omega _{+}-\frac{%
\omega }{2}\right) \right\} \,,  \notag \\
U_{2}& =-a^{1/2}\beta _{+}^{-1/2}\sin \frac{\theta }{2}\left\{ \sin \left(
\omega _{-}+\frac{\omega }{2}\right) +\beta \sin \left( \omega _{-}-\frac{%
\omega }{2}\right) \right\} \,,  \notag \\
U_{1}& =\phantom{-}a^{1/2}\beta _{+}^{-1/2}\sin \frac{\theta }{2}\left\{ \cos \left(
\omega _{-}+\frac{\omega }{2}\right) +\beta \cos \left( \omega _{-}-\frac{%
\omega }{2}\right) \right\} \,.
\end{align}%
where $\omega _{\pm }=\left( \phi \pm \psi \right) /2$.

\section{Decoupling the equation of motion}

\label{General formulas} For convenience the quaternion equation of motion
(\ref{regularized equation of motion}) can be decoupled into two complex
equations. Premultiplying (\ref{regularized equation of motion}) with $%
\left( -1-\beta ^{2}\right) \bar{\mathbb{U}}_{K}/\left( 2a\right) $ and also
postmultiplying it by $\bar{\mathbb{R}}$ while keeping only the $1$, $%
\mathfrak{i}$ part in $\mathcal{O}(\varepsilon )$ one obtains\cite{Vrbikbook}
\begin{align}
& -\mathfrak{i}\left( \beta _{-}-\beta z_{-}\right) \frac{%
a^{\prime }}{2a}+\mathfrak{i}z_{+} \beta ^{\prime }+4%
\mathfrak{i}\beta \beta _{+}^{-1}\beta ^{\prime }  \notag \\
& +\left( 2\beta _{-}+\beta z_{-} \right) Z_{3}-4\beta
_{+}s_{p}^{\prime }+(1+\beta z)\left( \mathfrak{\mathbb{D}}+8z\frac{d%
\mathfrak{\mathbb{D}}}{dz}+4z^{2}\frac{d^{2}\mathfrak{\mathbb{D}}}{dz^{2}}%
\right)  \notag \\
& +\left( 1+\beta z^{-1}\right) \mathfrak{\mathbb{D}}^{\ast }+(1-\beta
z)\left( \mathfrak{\mathbb{D}}+2z\frac{d\mathfrak{\mathbb{D}}}{dz}\right)
+\left( 1-\beta z^{-1}\right) \left( \mathfrak{\mathbb{D}}+2z\frac{d%
\mathfrak{\mathbb{D}}}{dz}\right) ^{\ast }=  \notag \\
& -\left( 1+\beta z^{-1}\right) (1+\beta z)^{2}\mathbb{Q\,},
\label{mozgegyenlet 1i 2}
\end{align}%
where $z_{\pm}=z \pm z^{-1}$ and $Z_{n}$ are the components of the angular velocity vector (\ref%
{angularvelocityKepler})

Similarly premultiplying equation (\ref{regularized equation of motion})
with $\left( 1+\beta ^{2}\right) \bar{\mathbb{U}}_{K}\mathfrak{k}/a$ and
then keeping only the complex part in $\mathcal{O}(\varepsilon )$ the second
complex equation is the following \cite{Vrbikbook}
\begin{eqnarray}
&&-8\frac{\beta _{+}\mathbb{S}}{\beta _{+}+\beta z_{+}}-8%
\frac{z_{-} \beta }{\beta _{+}+\beta z_{+}
}z\frac{d\mathbb{S}}{dz}+8z\frac{d\mathbb{S}}{dz}+8z^{2}\frac{d^{2}\mathbb{S}%
}{dz^{2}}  \notag \\
&&+Z_{1}\mathfrak{i}\beta _{-}\frac{2\beta _{+}z_{+} +\beta
\left( z^{2}+z^{-2}+6\right) }{\beta _{+}+\beta z_{+} }-8%
\mathfrak{i}\frac{\beta _{+}b}{\beta _{+}+\beta z_{+} }
\notag \\
&&+Z_{2}z_{-} \frac{\beta \beta _{+}z_{+}
+2\left( 1+\beta ^{4}\right) }{\beta _{+}+\beta z_{+} }=
-\left( 1+\beta z^{-1}\right) \left( 1+\beta z\right) \mathbb{W}\left(
z\right) .  \label{mozg jk}
\end{eqnarray}

\section{The solution for $\mathbb{D}$ and $\mathbb{S}$}

Similarly by pairing the powers of $z$ in the complex equations (\ref%
{mozgegyenlet 1i 2}) and (\ref{mozg jk}) two additional equation for $%
\mathbb{D}$ and $\mathbb{S}$ can be gained
\begin{eqnarray}
\mathbb{D} &\mathbb{=}&\mathbb{-}\frac{1}{4}\sum_{\substack{ n=-\infty  \\ %
n\neq -1,0}}^{n=\infty }\left[ \frac{\beta \left( n+\frac{1}{2}\right)
Q_{n-1}+\left( n-\frac{1}{2}\right) Q_{n}+\frac{1}{2}\overline{Q_{-n}}}{%
n^{2}\left( n+1\right) }\right.  \notag \\
&&\left. +\frac{\beta ^{2}\left( n+\frac{3}{2}\right) Q_{n}+\left( n+\frac{1%
}{2}\right) Q_{n+1}-\frac{1}{2}\beta ^{2}\overline{Q_{-n-2}}}{n\left(
n+1\right) ^{2}}-\frac{\frac{1}{2}\beta \overline{Q_{-n-1}}}{n^{2}\left(
n+1\right) ^{2}}\right] z^{n},  \label{kifejezesDre} \\
\mathbb{S} &\mathbb{=}&\mathbb{-}\frac{\mathfrak{i}}{4}Im\left[
\sum_{n=2}^{\infty }\left( \frac{\beta \mathcal{W}_{n-1}}{\left( n-1\right) n%
}+\frac{\beta _{+}\mathcal{W}_{n}}{n^{2}-1}+\frac{\beta \mathcal{W}_{n+1}}{%
n\left( n+1\right) }\right) z^{n}\right] .  \label{kifejezes Sre}
\end{eqnarray}

\section{The $\mathbb{D}$ and $\mathbb{S}$ quantity in case of the (post)$^1$-newtonian effect}
The complicated quantity $\mathbb{D}$ is given only up to second $\beta$ order  
\begin{align}
	\mathbb{D}= \frac{m \beta z}{2 a}(1-2 \eta ) +
	\frac{m \beta^2}{8 a z^2} \left(30 -2  z^4-9 \eta +5 z^4 \eta \right)+O(\beta^3)
\end{align}
while $\mathbb{S}$ is zero.

\section{Gravitational radiation: $\mathbb{D}$ and $\mathbb{S}$}
The fairly complicated $\mathbb{D}$ quantity is given in second $\beta$ order
\begin{align}
\mathbb{D}= -\frac{16}{15} \mathfrak{i} z \eta  \beta \left(\frac{m}{a}\right)^{5/2} +
\mathfrak{i}\frac{\eta  \beta ^2}{45 z^2} \left(\frac{m}{a}\right)^{5/2}\left(537+233 z^4\right)
+O(\beta^3)
\end{align}
while $\mathbb{S}$ is zero.

\end{document}